\documentclass{aa}
\usepackage[dvips]{graphics}
\usepackage{astron}

\begin{document}

\thesaurus{
06(03.20.2; 05.15.1; 08.02.1; 08.09.2: Haro 6-37;
08.16.5; 13.09.6)
}
\title{
Discovery of a close companion to the young star 
\object{Haro\,6-37}\thanks{Based on observations collected
at the Calar Alto Observatory (Spain),
operated by the German-Spanish Astronomical Center.
}
}

\author{
A. Richichi
\inst{1}
\and
R. K{\"o}hler
\inst{2}
\and
J. Woitas
\inst{3}
\and
Ch. Leinert
\inst{3}
}

\institute{
Osservatorio Astrofisico di Arcetri,
L.go Enrico Fermi 5, I--50125 Firenze, Italy
\and
Astrophysikalisches Institut Potsdam,
An der Sternwarte 16, D--14482 Potsdam, Germany
\and
Max-Planck-Institut f{\"u}r Astronomie,
K{\"o}nigstuhl \/17, D--69117 Heidelberg, Germany
}

%\offprints{A. Richichi}
%\mail{arichichi@arcetri.astro.it}
\offprints{A. Richichi \protect\\
              e-mail: \tt richichi@arcetri.astro.it}

\date{Received December 7, 1998 - Accepted January 22, 1999}
%\titlerunning{Discovery of a close companion to Haro\,6-37}
\maketitle

\begin{abstract}
We report lunar occultation (LO) and
speckle interferometry observations of \object{Haro\,6-37},
all carried out at $\lambda$=$2.2\mu$m.
The main finding is the discovery of a new close companion
by LO, subsequently confirmed by
speckle. The two techniques yield
values in excellent agreement, with
$0\farcs331$ and $181\degr$ for the separation and
position angle respectively. Additionally, the LO
data show evidence of what might be interpreted
as scattered light from the inner part of the dust disk
which is known to exist around \object{Haro\,6-37} from
mm-radio observations (Osterloh \& Beckwith
\cite{Osterloh}). However this last conclusion is subject
to some uncertainty because of possible contamination
of the signal during data acquisition. 
LO and speckle results are in good agreement
also for what concerns the brightness ratio of the
newly found companion to the main component,
R=0.10 or $\Delta$K=2.5.
We discuss the possible implications on the
general picture of \object{Haro\,6-37}, in
particular for what concerns the ages of the
components and the dust disk.
\end{abstract}

\section{Introduction}\label{intro}
Pre-main sequence (PMS) stars represent the link between the
early stages of star formation, and the beginning of life as a normal
star on the main sequence. This phase, lasting $\approx 10^5-10^7$
years, is strongly influenced by the initial conditions of the
surrounding environment, and characterizes in a significant
way the newborn stars in the remaining $\approx 10^6-10^{10}$ years
of their lifetimes.
Therefore, observations of PMS stars play an important
role in our understanding of the processes of star formation
and stellar evolution,
and as such they have been studied with increasing
interest at several wavelengths from the X-rays to the radio domains.

In particular, in the past decade it has become increasingly evident
that many PMS stars exhibit two important features, namely the presence
of circumstellar disks and a binary (or multiple) nature. The circumstellar
disks, first inferred from 
millimeter observations and later observed
in some cases also by direct imaging at visual/IR wavelengths, are thought
to be the remnant of the original parent condensation. The presence of 
one or more physically associated
companions, on the other hand, is the
signature of a common formation process and provides a key to
discriminate between the different possible
formation mechanism at play. The study of such binary or multiple systems
also allows us to make inferences on their evolution in the presence of the
complex environment of the parent star forming region (SFR), 
and once they reach the main sequence.
Of particular interest is the understanding of the interactions that
take place between the circumstellar (or sometimes circumbinary) disks,
and the binary components. 
This in turn has direct implications on the formation
of planetary systems, another topic which is receiving the 
highest attention in modern observational astronomy.
Useful references to many topics related to PMS stars research
can by found in the work by Hartmann \cite{hartmann};
recent compilations of PMS multiple stars  are given by
Ghez \cite{ghez98} and by Mathieu \cite{mathieu}.

In this line of research, high angular resolution observations in
the near-infrared (NIR) play an important role next to the
radio and visual ranges. In fact, most radio observations
lack the resolution to study the hot, inner parts of the circumstellar
disks and to reveal the closest binary components. Observations in the
visual range, on the other hand, are severely limited by extinction
due to both the circumstellar dust, and the local dust in the
surrounding SFR. The typical size of circumstellar disks 
and binary separations (10--100\,AU) are such, that even in the
nearest SFRs with a high number of PMS stars (Taurus-Auriga and
Scorpius-Ophiuchus, both at $\approx 150$\,pc), 
the natural limit in angular resolution
set by atmospheric turbulence will limit significantly our ability
to investigate their properties. As a consequence, this field of
research has seen an increasingly important contribution from
high angular resolution techniques such as
lunar occultations (e.g. 
Leinert et al. \cite{lei_dgtau},
Richichi et al. \cite{rich_binatau},
Simon et al. \cite{simon_survey})
and speckle interferometry
(e.g. Ghez et al. \cite{ghez}, Leinert et al. \cite{lei_survey}). 
More recently,
also other methods have been successfully employed
(e.g. adaptive optics Roddier et al. \cite{roddier}, Close et al. \cite{close};
speckle holography Petr et al. \cite{petr}).
Among the main results to be mentioned, is the fact that the frequency
of binary systems among PMS stars appears to be significantly higher
than for main sequence stars, at least within the limitations 
of separation range and masses and in the
number of stars, which are intrinsic
in the currently available surveys. This is true at
least for the Taurus-Auriga SFR, to which \object{Haro\,6-37} belongs.
A detailed discussion
was given by Simon et al. \cite{simon_survey}. 
However our conclusions are still limited by the available
statistics:
in this sense, it is important to increase the numbers of
observed binary and multiple systems.

With this work, we report a newly discovered companion around
\object{Haro\,6-37}, which we detected initially by a lunar occultation (LO)
event, and subsequently confirmed by speckle interferometry observations.

\section{Observations}\label{observations}
\begin{figure}
\resizebox{\hsize}{!}{\includegraphics{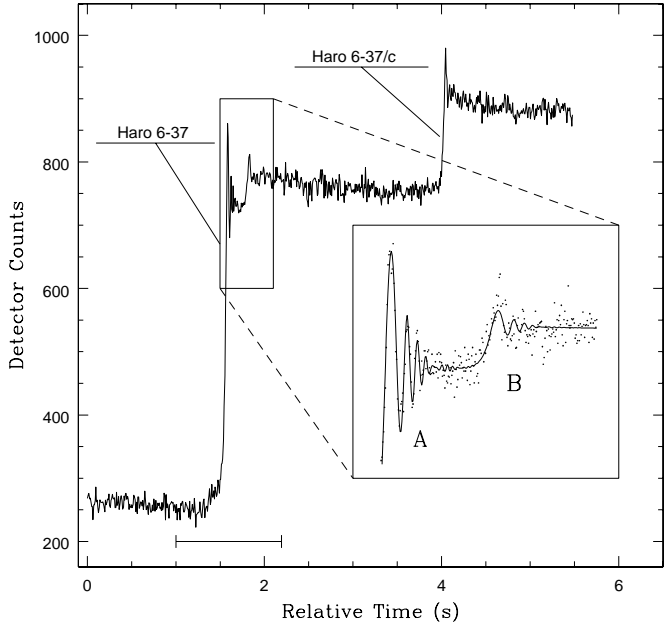}}
\caption
{
The larger figure shows a long portion of the occultation trace, 
including the
distant {\sl c} component. These data are rebinned
by a factor of 4 to improve the presentation.
The inset is an enlargement to show
the detection of the newly discovered companion.
The occultation data are shown as dots and our fit by a 
multiple star model as a solid line. The horizontal bar
on the bottom marks the portion detailed in Fig.~\ref{fig_disk}.
}\label{fig_occ}
\end{figure}
\begin{figure}
\resizebox{\hsize}{!}{\includegraphics{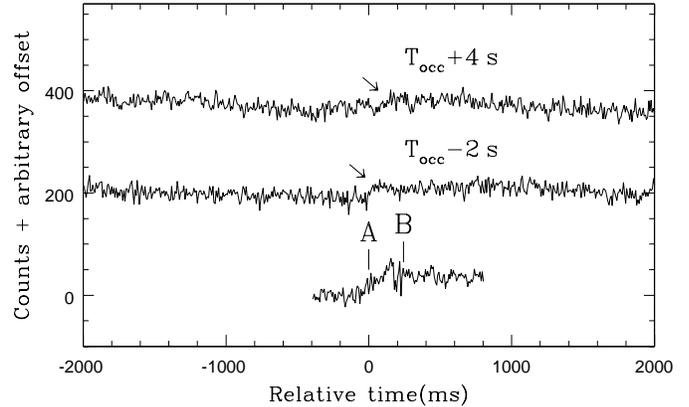}}
\caption
{
Comparison of ``glitches'' in relatively distant
portions of the lightcurve (marked by the arrows in the
two top curves), with the possible signature of
extended emission close to the main component (lower
curve). This latter is shown after subtraction of
the fringe patterns of the A and B components, whose
position is marked.
Data rebinned by a factor of 4.
}\label{fig_disk}
\end{figure}
An occultation of \object{Haro\,6-37} was observed on November 16, 1997
at the Calar Alto 3.5\,m telescope equipped with an IR
fast photometer. This instrument, 
based on an InSb detector cooled to solid nitrogen temperature,
is usually devoted to
slit--scanning speckle interferometry, but thanks
to its fast time response and large data storage capabilities,
it has been successfully employed also
to observe LO events, in particular of
T Tauri stars (e.g. Leinert et al. \cite{lei_dgtau},
Richichi et al. \cite{rich_binatau}).

The LO data were
recorded with a sampling of 1.95\,ms in
a standard wide band K filter
($\lambda_{\rm 0}=2.36\,\mu$m, $\Delta \lambda=0.46\,\mu$m).
It is shown in Fig.~\ref{fig_occ}, together with our best fit.
The fit has been obtained using a least--squares method
which is described in Richichi et al. \cite{rich_lsm}. This
approach is well suited to derive the parameters of a simple
source model (such as a binary or multiple star). However,
when an extended, geometrically complex component is present,
a model--independent analysis is necessary. For this, we use
an iterative method based on the Lucy's deconvolution algorithm,
and expecially designed for LO data (the so-called CAL algorithm,
Richichi \cite{rich_cal}). 

The speckle observations were carried out also at the 3.5\,m telescope
on Calar Alto shortly after the LO event
on November 24, 1997.
We used MAGIC, a 256 x 256 pixel NICMOS3 camera (Herbst et al. 
\cite{MAGIC})
in its high-resolution configuration at the f/45 focus, also
in the K band.  Pixel scale and orientation 
were determined by
astrometric fits to images of the Trapezium cluster, yielding
$(0\farcs0710\pm0\farcs0002)/\mathrm{pixel}$ and 
$-0\fdg28\pm0\fdg05$ respectively.
A sequence of 4 groups of 250 frames were recorded on \object{Haro\,6-37},
each with an integration time of 0.8\,s. These groups of
fast integrations were alternated with a similar sequence on
a nearby point--like reference star, SAO~94117. During data reduction
this latter appeared to be itself a binary star, and thus not
suitable for calibration. Instead, we used
as a reference HIC~19797 (HIC = Hipparcos Input Catalogue), which
was similarly observed in the course of the same observational
program shortly before \object{Haro\,6-37}. In spite of an
angular distance of $\approx 8\degr$, the calibration 
by this star was satisfactory.

\begin{figure}
\resizebox{\hsize}{!}{\includegraphics{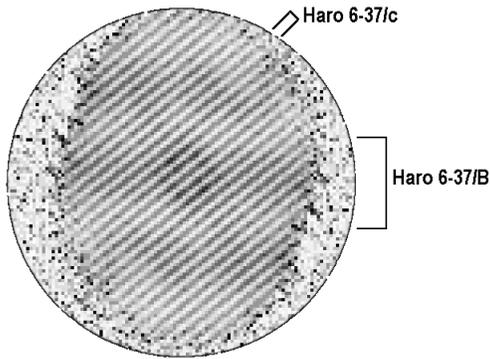}}
\caption
{The modulus of the 
complex visibility of \object{Haro\,6-37},
reconstructed from our speckle data. 
The frequency spacing and orientation of the
fringe systems of the two companions are highlighted.
The circle is a portion of the Fourier space
with a radius of 6.2 arcsec$^{-1}$.
}\label{vismod}
\end{figure}

The speckle data were processed in order to derive the
modulus of the complex visibility (i.$\,$e.\ the Fourier transform
of the object brightness distribution) from power
spectrum analysis (see Fig.~\ref{vismod}). 
The phase was computed using the 
Knox-Thompson \cite{KnoxThomp74} algorithm, as well as the
bispectrum method (Lohmann et al. \cite{Lohmann83}).
The modulus and phase were then fitted by a triple star model
to derive the brightness ratio, separation and position
angle of the components.  Fits to different subsets of the data give an
estimate for the standard deviation of the parameters.

\section{Results and discussion}\label{results}
The occultation data shown in Fig.~\ref{fig_occ} reveal
the presence of two companions to \object{Haro\,6-37}.
One of them is the relatively distant component 
firstly discovered
by Moneti \& Zinnecker \cite{moneti_zin} by imaging,
and dubbed \object{Haro\,6-37/c}. 
Its projected angular distance recovered from the LO data is 
$\approx$$0\farcs81$, but there is some uncertainty because
the occultation rate of this component appears to differ
significantly from that of the primary. This is due to
the fact that the true separation is quite large (see below).
Taking into account the occultation geometry,  it turns out that
\object{Haro\,6-37/c} was occulted by a point on the lunar
limb 4.4\,km away from the primary; this accounts for a
significant difference in the local slope of the lunar limb, estimated 
in about 10$\degr$.  At the
same time, the distance between the two components is comparable to one-half
of the diaphragm (6$\arcsec$) used in our LO measurement, making
our LO-based estimated of the
brightness ratio, R$\approx$0.28, subject to
the possible uncertainties of non-uniform
detector response across the field. In summary, the relatively
large angular distance between these two objects makes it a
difficult target for LO. However, \object{Haro\,6-37/c} is
easily resolved in our speckle data, as shown
in Fig.~\ref{vismod}. These yield a true position angle
PA=$38\fdg6\pm0\fdg1$ and separation $\rho$=$2\farcs627\pm0\farcs008$,
with a brightness ratio R=0.35$\pm$0.01 in K. 
We note that the LO values are consistent within the 
uncertainties above mentioned. \object{Haro\,6-37/c} has been 
relatively well studied already in the literature,
and we will not concern ourselves further with it.

The second, closer component is more interesting, since it 
represents a new detection. Our fit to the LO data, also
shown in Fig.~\ref{fig_occ}, yields a projected separation 
for the A-B pair
$\rho_{\rm p}$=$0\farcs0857\pm0\farcs0011$ along
PA$_{\rm LO}$=$285\fdg6\pm0\fdg2$, with a brightness ratio 
R=$0.112\pm0.001$ in K. 
In this case, the vicinity to the primary ensures a very
reliable measurement.
A model fit to the speckle data with the inclusion
of this third component in the system, led to the values
PA=$180\fdg7\pm0\fdg9$, $\rho$=$0\farcs331\pm0\farcs005$ 
and R=$0.10\pm0.02$, also in K.
It can be noted that the two 
independent measurements by different techniques
are highly consistent, particularly for what concerns
the separation. The brightness ratio is slightly discrepant,
but we must add that the LO data seem to show the
presence of additional extended emission close
to the location of \object{Haro\,6-37/A}.

This was isolated from the fringe patterns of the A
and B component and analyzed by the CAL algorithm.
The result is consistent with an extended emission
of roughly gaussian shape, 
having a projected FWHM=$0\farcs075$ and traced out
to a total extent of about $0\farcs2$.
This could represent the signature of the
inner parts of the dust disk detected by mm-radio
observations, but at the same time we must also point out that 
a few random variations of the background, or
DC-offset in the amplification scheme, are also present in the data. 
Fig.~\ref{fig_disk} shows some of these variations,
compared to the signal which is being tentatively 
attributed to the extended emission. Although it
can be noted that the intensity is at least $\approx 3$
times larger for this latter, the order of magnitude and the
typical timescales are comparable. Unfortunately,
this prevents us to make a definitive conclusion.

One consequence is that also the brightness
ratio derived from the LO event must be revised.
If the integrated intensity of
suspected extended emission (be its nature real
or not) is subtracted from that of the A component, the
brightness ratios are 1:0.124:0.101 for the
A, B and extended components respectively.
This brings the LO and speckle determinations
for the A-B brightness ration in very good agreement.
We also note that such extended emission would
be extremely difficult to detect in the
speckle observation, causing a slight, slow
decrease in the visibility, which we cannot confirm
in our data given the available SNR. The brightness
ratio of the A-B pair would not be affected,
to a first approximation, by the neglection in the fit of such
an extended component.

On 9 January 1998, shortly after our LO and speckle observations,
K-band photometry of \object{Haro\,6-37} was obtained using the NIR
camera IRCAM3 at the UKIRT telescope. 
The derived magnitude were K=$(7.92\pm0.08)$ mag
and K=$(8.97\pm0.08)$ mag for the central star and the
distant companion. From the flux ratio derived above,
this leads to K=$(8.03\pm0.09)$ mag for the A+B component,
and K=$(10.48\pm0.09)$ mag for the newly detected companion.
For the purpose of the discussion to follow, it is
important to establish whether any variability is present.
This is a common feature among T Tauri stars, although
it is recognized that its magnitude is more important
in the visual than in the near-IR.
Hartigan et al. \cite{hartigan} find that variability 
causes a mean error
of 0.17\,mag in the J-band for one component of a binary system.
In our case, we can compare our values to the photometry
obtained by Moneti \& Zinnecker \cite{moneti_zin}
almost exactly ten years before, giving
K=7.71\,mag and 8.58\,mag for the A+B component and the 
distant companion, respectively. A decrease of about
0.2\,mag is apparent for the A+B component over
this time period.

The detection of a new companion is hardly surprising, since it
is now well established that young low-mass stars show a high
incidence of binary frequency, and particularly so in
the Taurus-Auriga SFR as was noted before. Nevertheless this
new result has some additional implications, since it makes
\object{Haro\,6-37} a triple system and because of the
presence of a circumstellar disk around it, established
by mm-radio observations. In particular, Osterloh \& Beckwith
\cite{Osterloh} derive a total disk mass of 0.017\,M$_{\sun}$,
a characteristic temperature at 1\,AU of 130\,K, and speak of
this system as an ``example of a consistent picture of a star
and a nearly standard reprocessing disk ({\it q}=0.65)''.

It is therefore interesting to look into the implications
of the newly detected companion for the general picture of
this system. It has been pointed out by Simon et al. \cite{simon_ages},
that our estimate of the ages of T Tauri stars can be biased
up to a factor of 2 by the presence of an undetected companion.
Hartigan et al. \cite{hartigan} have placed \object{Haro\,6-37} and
\object{Haro\,6-37/c} into the HR-diagram using 
the J-band magnitude as luminosity indicator. 
By comparison with theoretical PMS tracks by D'Antona \& Mazzitelli
\cite{dm94} and Swenson \cite{swenson}, 
they find both stars to be coeval with
an age of about $6\times10^5\,\mathrm{yr}$. 
The detection of a third component within this system
could change this conclusion, only if it had
a very red color.
In fact, if we assume that the A-B flux ratio is constant
between J and K, the J-band brightness of \object{Haro\,6-37/A} would
be overestimated by about 0.10\,mag only. This is not
significant, given that 
Hartigan et al. \cite{hartigan} estimate that
variability and interstellar extinction already produce
a 0.19\,mag uncertainty in the HR-diagram.
Also the spectral type of the main component is essentially
unaffected, because of the low flux ratio.
However, we note that in this regime of masses and ages,
the PMS tracks are almost horizontal in
the HR-diagram, so that if the color of \object{Haro\,6-37/B}
was significantly redder, this would imply a reduced
mass and a larger age for the primary star, and in
turn a significant age difference between this latter and the more
distant companion.
In any case the interesting question remains, if
also the newly detected companion is coeval with the two other components.
This can be answered when additional
spectra and J-band photometry become available.

For what concerns the disk, the new companion does
not seem to alter significantly the present understanding
of the system. The values derived by Osterloh \& Beckwith \cite{Osterloh}
indicate a compact dust distribution, with an inner
dust formation radius of only 0.013\,AU, and a $\tau_{\rm 1.3mm}$=1
radius of 15\,AU. By comparison, the newly detected companion
is at a distance of $\approx$46\,AU, and this should be considered
as a minimum value for the separation in the absence of any
orbit determination. Therefore, no real influence on the inner
dust disk is to be expected, apart from tidal forces or a clearing
of the outer dust, effects which at present are beyond our 
observational capabilities.

We already mentioned that 
the additional signal apparent in the LO data,
if real, would imply an extended envelope with
a projected FWHM=$0\farcs075$ or about 10\,AU
and about 10\% of the intensity of the primary star.
The most natural explanation would be that this
is scattered light from the above mentioned dust disk,
and we note that already in other cases similar
detections were obtained by LO. For instance,
Leinert et al. \cite{lei_dgtau} reported an extended
emission attributed to scattered light around
DG~Tau, with FWHM of about 7\,AU
and 25\% of the stellar brightness in K.
Additionally, we note that the CAL analysis places
this emission approximately in the middle 
of the A and B components, with the two stars
at the edges. This would suggest
a circumbinary disk, but the picture is necessarily
limited because of the projection effects
intrinsic in the LO method.

In conclusion,  the newly detected companion does not alter
significantly our understanding of the \object{Haro\,6-37}
system, but it would be interesting to assess more
in detail the relative spectral energy distribution of
the A and B components, and to obtain a spectrum
of this latter. This is at the limit of current
observational possibilities, but would be a necessary
step to determine whether
also \object{Haro\,6-37/B} is a classical T Tauri star.
This would confirm the result of Prato \& Simon \cite{prato}
who have found that in 12 young binary systems with classical 
T Tauri behaviour,
{\it both} components are CTTS. 
Additionally, direct imaging at very high angular resolution
of the A and B components could reveal the (circumbinary)
disk in scattered light.
\object{Haro\,6-37/B} appears to be  ideally suited
for observations by the new generation of large
ground-based imaging interferometers.

{\sl Note added in proof}. After submission of the present work, we became
aware of a paper by G. Duch{\^e}ne \cite{duchene}, in which the author
quotes the discovery of the binarity of the main component of
\object{Haro\,6-37}, observed with adaptive optics by another
group (Monin et al., in preparation). The reported result, obtained just one
month after our detection, is in very good agreement with the values
given in the present work for what concerns angular separation and
position angle. Some disagreement with our determination
seems to exist for the brightness ratio,
which Duch{\^e}ne reports as $\Delta$K=1.57$\pm0.05$.

\acknowledgements
This research has made use of the {\it Simbad} database, operated at CDS,
Strasbourg (France). We thank P. Kalas for obtaining the UKIRT photometry.

\bibliographystyle{astron}
\bibliography{mnemonic,example}
{}
%\listofobjects

\end{document}